\documentclass[aps,twocolumn,showpacs,preprintnumbers,amsmath,amssymb,floatfix,nofootinbib]{revtex4-1}
\usepackage{graphicx}

\usepackage{color}
\usepackage{colordvi}

\usepackage[normalem]{ulem}
\usepackage{slashed}

\newcommand{\C}{{\cal C}}
\usepackage{color}
\newcommand{\beqn}{\begin{eqnarray}}
\newcommand{\eeqn}{\end{eqnarray}}
\newcommand{\be}{\begin{equation}}
\newcommand{\ee}{\end{equation}}

\newcommand{\ba}{\begin{array}{c}}
\newcommand{\bat}{\begin{array}{cc}}
\newcommand{\ea}{\end{array}}

\newcommand{\bi}{\begin{itemize}}
\newcommand{\ei}{\end{itemize}}

\newcommand{\invfb}{fb$^{-1}$}
\usepackage{pifont}%
\newcommand{\cmark}{\ding{51}}%
\newcommand{\xmark}{\ding{55}}%

\definecolor{sexyred}{RGB}{220,20,60}

\newcommand{\cO}{{\cal O}}

\newcommand{\gsim}{\stackrel{>}{_\sim}}

\usepackage{xcolor}

\begin{document}

\preprint{Imperial-TP-2018-JECM-01}
\preprint{LMU-ASC 18/18}


\title{Anomalies in Bottom from new physics in Top}

\author{Jos\'e Eliel Camargo-Molina${}^{\,\circ}$, Alejandro Celis${}^{\,\circ\circ}$, Darius A. Faroughy${}^{\,\circ\circ\circ}$ \vspace{0.3cm}}

\affiliation{${}^{\circ}$ Department of Physics, Imperial College London,\\ South Kensington campus, London SW7 2AZ, UK.
}
   \email{J.camargo-molina@imperial.ac.uk}
   
   \affiliation{${}^{\circ\circ}$ Ludwig-Maximilians-Universit\"at M\"unchen, 
   Fakult\"at f\"ur Physik,\\
   Arnold Sommerfeld Center for Theoretical Physics, 
   80333 M\"unchen, Germany   }
   
   \email{Alejandro.Celis@physik.uni-muenchen.de}
   
   \affiliation{${}^{\circ\circ\circ}$ J. Stefan Institute, Jamova 39, P. O. Box 3000, 1001 Ljubljana, Slovenia}

\email{darius.faroughy@ijs.si}

\begin{abstract}

We analyze the possibility to accommodate current $b \to s \ell^+ \ell^-$ anomalies with TeV-scale mediators that couple to right-handed top quarks and muons, contributing to $b \to s \ell^+ \ell^-\,$ at the one-loop level. We use the Standard Model Effective Field Theory (SMEFT) framework but also look at specific scenarios by taking into account all possible irreducible representations of the Lorentz and Standard Model gauge group for the mediators.    From a global fit of $b \to s \ell^+ \ell^-$ data and LEP-I observables we find that the Wilson coefficients of two SMEFT operators: $\mathcal{O}_{\ell u}= (\bar \ell_{ L \mu} \gamma^{\alpha}  \ell_{ L \mu}) (\bar t_R \gamma_{\alpha}  t_R  )$ and  $\mathcal{O}_{e u}=  (\bar \mu_{R} \gamma^{\alpha}  \mu_{R})  (\bar t_R \gamma_{\alpha}  t_R  )$ need to satisfy $\C_{eu} \sim \C_{\ell u}$. New physics enters then in $b \to s \ell^+ \ell^-$ mainly through the operator $\mathcal{O}_9 = (   \bar s  \gamma_{\mu}   P_{L}   b   )( \bar \ell \gamma^{\mu} \ell )$ of the Weak Effective Theory.  After discussing all possible mediators, we concentrate on two scenarios:   A vector boson in the irreducible representation $Z_{\mu}^{\prime} \sim (1,1,0)$ of the Standard Model gauge group with vectorial coupling to muons, and a combination of two leptoquarks: the scalar $R_2 \sim (3,2,7/6)$ and the vector $\widetilde U_{1 \mu} \sim (3,1,5/3)$. We derive LHC constraints by recasting di-muon resonance, $pp \to t \bar t t \bar t$ and SUSY searches.  Additionally, we analyze the prospects for discovering these mediators during the high-luminosity phase of the LHC.   
\end{abstract}

\maketitle

%

\section{Introduction}

The observed pattern of deviations from the Standard Model (SM) in $b \to s \ell^+ \ell^-$ transitions~\cite{Aaij:2014ora,Aaij:2017vbb,Aaij:2013qta,Aaij:2015oid,Wehle:2016yoi} suggests the presence of new physics (NP) that violates lepton flavour universality.     Global analyses of the experimental measurements within the Weak Effective Theory find that only a few effective operators are needed to consistently explain  the observed deviations from the SM~\cite{Capdevila:2017bsm,Altmannshofer:2017yso,Geng:2017svp,Ciuchini:2017mik,DAmico:2017mtc,Hiller:2017bzc}.\footnote{One exception is the current measurements of $\Lambda_b \to \Lambda \ell^+ \ell^-$, which show some tension with the measurements in semileptonic $B$ decays~\cite{Meinel:2016grj}.}        

A NP scenario that has been explored recently in this context is that of a mediator that couples predominantly to right-handed up-type quarks and to muons.   The required contributions to explain the $b \to s \ell^+ \ell^-$ anomalies arise at the one-loop level in this case.\footnote{Departures from the SM have also been observed in $b \to c \tau \nu$ transitions.  The scenario proposed here cannot address these anomalies, so one would need to extend this dynamical setting in order to accommodate them.  See~\cite{Ciezarek:2017yzh} for a review of the current experimental situation.  }  
   This idea has been presented in terms of the Standard Model Effective Field Theory (SMEFT)~\cite{Celis:2017doq}, as well as with specific models: a scalar leptoquark~\cite{Becirevic:2017jtw}, and a top-philic $Z^{\prime}$ boson~\cite{Kamenik:2017tnu,Fox:2018ldq}.    Given the large value of the top-quark mass and the structure of the Cabibbo--Kobayashi--Maskawa (CKM) matrix, the largest effects in $b \to s$ transitions will be generated when the mediator couples to top-quarks.   We therefore focus on this case.

In this work we analyze this scenario within the EFT framework as well as with particular models, considering all possible mediators that give rise to tree-level matching contributions to the SMEFT operators on which we are interested.    We consider constraints from LEP-I on the modifications of the $Z$ properties, including the necessary one-loop matching corrections at the electroweak (EW) scale.    We also discuss the constraints from high-$p_T$ searches at the LHC and analyze future prospects for the high-luminosity phase of the LHC.

The new findings in this letter are:
\begin{itemize}
\item  We find that the SMEFT operators
\begin{align} \label{twoop}
[\mathcal{O}_{\ell u}]_{\mu \mu tt} &= (\bar \ell_{ L \mu} \gamma^{\alpha}  \ell_{ L \mu}) (\bar t_R \gamma_{\alpha}  t_R  )  \,, \nonumber \\
[\mathcal{O}_{eu}]_{\mu \mu tt} &= (\bar \mu_{R} \gamma^{\alpha}  \mu_{R}) (\bar t_R  \gamma_{\alpha}  t_R  )   \,,
\end{align}
can accommodate the $b \to s \ell^+ \ell^-$ data and the constraints from LEP-I measurements when their corresponding Wilson coefficients satisfy $\C_{\ell u} \sim \C_{eu}$.     For $\Lambda \sim 1$~TeV, we find $\C_{\ell u} \sim \C_{e u} \sim -1.7$.
\item We explore all the possible mediators that can generate the required NP pattern ($\C_{\ell u} \sim \C_{e u}$ with negative values).     We find that, among the colorless mediators a minimal scenario consists of having a vector boson in the irreducible representation $Z_{\mu}^{\prime} \sim (1,1,0)$ of the SM gauge group $\mathrm{SU(3)}_C \times \mathrm{SU(2)}_L \times \mathrm{U(1)}_Y$ with vectorial coupling to muons.  For the mediators carrying color,  we find a viable scenario with a combination of two leptoquarks, the scalar $R_2 \sim (3,2,7/6)$ and the vector $\widetilde U_{1 \alpha} \sim (3,1,5/3)$.
\item By recasting different high-$p_T$ searches at the LHC we find that the LHC is already probing the interesting region of TeV masses for these mediators.
\end{itemize}

This article is organized as follows.   The EFT framework used is discussed in Sec.~\ref{seceft}.   In Sec.~\ref{seclowph} we present the phenomenological analysis of low energy observables, including those coming from flavour physics and LEP.   Possible mediators that can accommodate the data are presented in Sec.~\ref{secmodels}.   Constraints from high-$p_T$ searches at the LHC are discussed in Sec.~\ref{sechighpt}.  Sec.~\ref{secenp} contains a small discussion of the possibility of NP in the electron channel.    Sec.~\ref{secdisc}  is a critical discussion of our results and comparison with previous related works. We conclude in Sec.~\ref{seccon}.  Appendix~\ref{appendixA} contains details about the LHC constraints on the leptoquarks.

\section{Effective Field Theory} \label{seceft}

\subsection{Standard Model Effective Field Theory}   \label{sec:smeft}  

When the NP particles are much heavier than the EW scale we can parametrize their effects at low energies via the SMEFT~\cite{Buchmuller:1985jz}.\footnote{In scenarios of strongly coupled dynamics behind electroweak symmetry breaking, with the Higgs arising as a pseudo-Goldstone boson, the nonlinear effective theory provides a more suitable low energy description~\cite{Buchalla:2013eza,Pich:2018ltt}.     }  Integrating out the heavy particles gives rise to a tower of effective operators suppressed by the mass of these particles, assumed here to be a common scale and denoted by $\Lambda$.  The dominant NP effects in the EFT power counting are encoded in operators of canonical dimension six\footnote{At dimension five there is only one operator, which provides neutrinos with a Majorana mass term after electroweak symmetry breaking.}
\begin{align}
\mathcal{L}_{\rm{SMEFT}}  = \mathcal{L}_{\rm{SM}}  +  \frac{1}{\Lambda^2}  \sum_i \C_i \mathcal{O}_i + \cdots
\end{align}
We adopt the non-redundant basis for the dimension six operators defined in~\cite{Grzadkowski:2010es}, known as the Warsaw basis.     

In the weak basis where the up-type quark and charged lepton mass matrices are diagonal, we consider the two operators in Eq.~\eqref{twoop} involving right-handed top quarks and muons (and its associated neutrino field).

\subsection{Weak effective Theory}

\begin{figure}[htp]
   \vspace{-1.4cm}
\begin{center}{
\includegraphics[width=4cm]{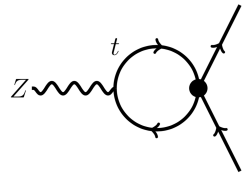}
  \vspace{-1.2cm}
\caption{\textit{One loop correction to $Z \to \mu^+ \mu^-$.}} \label{figZ}}
\end{center}
\end{figure}

After electroweak symmetry breaking (EWSB), the operators in~\eqref{twoop} modify the $Z$ boson couplings to the muons at the quantum level, see Figure~\ref{figZ}.   This is due to the operators $\mathcal{O}_{\ell u}$ and $\mathcal{O}_{eu}$ mixing under renormalization group evolution with $(\varphi^{\dag}  i  \overleftrightarrow D_{\mu} \varphi ) (\bar \ell \gamma^{\mu}  \ell)$ and $(\varphi^{\dag}  i  \overleftrightarrow D_{\mu} \varphi ) (\bar e \gamma^{\mu}  e)$ of the Warsaw basis~\cite{Jenkins:2013wua}.\footnote{In the conventions used $\varphi^{\dag}  i  \overleftrightarrow D_{\mu} \varphi = \frac{g_Z}{2} v^2  Z_{\mu} + \cdots$ where $v \simeq 246$~GeV and $\varphi$ is the SM Higgs doublet.}

We can parametrize these effects by     
\begin{align}  \label{Zlag}
\mathcal{L} = \frac{g_Z}{2}  \bar \mu    \gamma_{\alpha}   \left(    \delta g_L  P_L    +    \delta g_R    P_R   \right) \mu Z^{\alpha}   \,,
\end{align}
with $g_Z = g/c_W$.   Taking into account the first leading logarithm from renormalization group evolution together with the finite parts of the one-loop correction we obtain\footnote{Using \texttt{DsixTools}~\cite{Celis:2017hod}, we verify numerically that keeping the first leading logarithm is a good approximation for $\Lambda \sim$~TeV.}
\begin{align}
\delta g_L  &=  -  \frac{3   y_t^2 }{8  \pi^2  }       \frac{v^2}{\Lambda^2}     \left[  \mathrm{log} \left( \frac{m_t}{\Lambda} \right)      -  \frac{4 s_{\theta_W}^2}{9}  (F +1)  +  \frac{F}{2}              \right] \C_{\ell u} \,, \nonumber \\
\delta g_R   &=  -  \frac{3   y_t^2 }{8  \pi^2  }       \frac{v^2}{\Lambda^2} \left[   \mathrm{log} \left( \frac{m_t}{\Lambda} \right)  -  \frac{4 s_{\theta_W}^2}{9}  (F +1)  +  \frac{F}{2} \right]  \C_{e u} \,,
\end{align}
with $y_t = \sqrt{2} m_t/v$ and $v \simeq 246$~GeV the vacuum expectation value of the Higgs field.  We have retained terms that are enhanced by the top-quark Yukawa.   We use the notation $s_\alpha \equiv \sin \alpha$, $\theta_W$ is the weak angle. The Wilson coefficients $\C_{\ell u}$ and $\C_{eu}$ are evaluated at the scale $\Lambda$.   We have dropped the flavour indices on the Wilson coefficients $[\C]_{\mu \mu tt}$ for simplicity.    The loop function $F$ is given by ($\tau_t = 4 m_t^2/M_Z^2$)
\begin{align}
F  =  -2  + 2 \sqrt{\tau_t -1} \arctan \left(  \frac{1}{\sqrt{\tau_t -1}}\right)    \,.
\end{align}
The inclusion of the one-loop matching corrections cancels the scale dependence of the leading renormalization group contribution.  We also verified the corresponding entry of the anomalous dimension matrix calculated in~Ref.~\cite{Jenkins:2013wua}.   Note that one-loop finite corrections could also originate from a UV completion of our framework.   We assume that these model dependent finite corrections are subdominant, similar assumptions have been made for instance in~\cite{Feruglio:2017rjo}.

\begin{figure}[h]
\begin{center}{
\includegraphics[width=4cm]{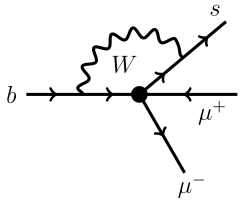}
\caption{\textit{One loop contribution in unitary gauge to $b \to s \mu^+ \mu^-$ from a four-fermion operator involving top quarks and muons.}\label{figintro}} }
\end{center}
\end{figure}

Below the EW scale, the top-quark is integrated out together with the $W,Z$ and the Higgs.  The operators in \eqref{twoop} give a one-loop matching contribution via the diagram shown in Figure~\ref{figintro} to the semileptonic operators 
\begin{align}  \label{o910}
O_{9}^{ij, \ell} &=  [   \bar d_i  \gamma_{\mu}   P_{L}   d_j   ][ \bar \ell \gamma^{\mu} \ell]   \,,  \quad O_{10}^{ij, \ell} &=  [   \bar d_i  \gamma_{\mu}   P_{L}   d_j   ][ \bar \ell \gamma^{\mu} \gamma_5 \ell] \,,
\end{align}
belonging to the weak effective Hamiltonian for $d_i \to d_j \ell^+ \ell^-$ transitions
\begin{align} 
\mathcal{H}_{\rm{eff}} = -  \frac{4 G_F}{\sqrt{2}}   \frac{\alpha_e}{4 \pi}  V_{ti}^*  V_{tj} \sum_{a}  \sum_{\ell}   \C_{a}^{ij,\ell}  O_{a}^{ij,\ell}     + \mathrm{h.c.} 
\end{align}
Here $G_F$ is the Fermi constant and $\alpha_e$ represents the fine-structure constant.

The leading contribution due to renormalization group evolution can be obtained using the one-loop anomalous dimension matrix obtained in~\cite{Jenkins:2013wua}.  The finite parts from the one-loop correction were calculated in~\cite{Aebischer:2015fzz}.  Keeping top-Yukawa enhanced contributions, the final results read~\cite{Celis:2017doq}

\begin{align} \label{myeqf}
\C_{9}^{ij,\mu}  &\simeq  \frac{x_t  v^2}{8 s_{ \theta_W}^2  \Lambda^2}    \left[    \log \left(  \frac{\Lambda}{M_W}\right)     + I_0(x_t)          \right]  (   \C_{\ell u} +   \C_{eu}  )  \,, \nonumber \\[0.2cm]
\C_{10}^{ij,\mu}   &\simeq   \frac{- x_t v^2}{8 s_{\theta_W}^2 \Lambda^2}   \left[    \log \left(  \frac{\Lambda}{M_W}\right)     + I_0(x_t)          \right]  (   \C_{\ell u} -   \C_{eu} )  \,,
\end{align}
with $x_t = m_t^2/M_W^2$ and $I_0(x_t) \simeq -0.71$ representing a loop function as in~\cite{Aebischer:2015fzz}; $\C_{\ell u}$ and $\C_{eu}$ are evaluated at the scale $\Lambda$. The Wilson coefficients $\C_{9,10}^{ij,\mu}$ are proportional to $m_t^2$ due to the required chirality flip in both quark legs.        We verified that one-loop matching corrections cancel the scale dependence of the leading renormalization group contribution in \eqref{myeqf}.

\section{Low energy phenomenology} \label{seclowph}

Modifications of the $Z \to \mu^+ \mu^-$ decay rate are constrained by lepton flavour universality tests in $Z$ decays performed at LEP-I.    We use the following measurement (see Sec.~7.2.1 in~\cite{ALEPH:2005ab})
\begin{align}    \label{zdec}
\frac{\Gamma_{\mu \mu}}{ \Gamma_{ee} }  = 1.0009 \pm 0.0028 \,, \qquad  \frac{\Gamma_{\tau \tau}}{ \Gamma_{ee} }  = 1.0019 \pm 0.0032 \,,
\end{align}
with a correlation $\rho = 0.63$.  The notation $\Gamma_{ff} = \Gamma(Z \to f^+ f^-)$ has been used.

The partial decay width for $Z \to \mu^+ \mu^-$ taking into account \eqref{Zlag} is given to linear order in the NP contributions by
\begin{align}  \label{propeq}
\frac{\Gamma(Z \to \mu^+ \mu^-)}{\Gamma(Z \to e^+ e^-)}  = 1 +  \frac{    c_{2 \theta_W}  \delta g_L    -  2  s_{\theta_W}^2 \delta g_R    }{A  }  \,,
 \end{align}
where $A \equiv 1  +  c_{4 \theta_W}/2-  c_{2 \theta_W} $ and  $c_\alpha \equiv \cos \alpha$.      

Modifications of the $Z$ coupling to leptons are also constrained by the leptonic asymmetry parameter determined at LEP-I. We use the measurement (see Table 7.4 in~\cite{ALEPH:2005ab})
\begin{align}  \label{Amuexp}
{\cal{A}}_{\mu} = 0.1456 \pm 0.0091 \,.
\end{align}
The leptonic asymmetry parameter is defined by
\begin{align}
&{\cal{A}}_{\mu}  =   \frac{   \Gamma(Z \to \mu_L^+ \mu_L^- ) -       \Gamma(Z \to \mu_R^+ \mu_R^- )  }{   \Gamma(Z \to \mu^+ \mu^- )   }  \, \nonumber  \\
 &= \frac{B}{A}  +  \frac{ c_{2 \theta_W} \delta g_L    + 2 s_{\theta_W}^2  \delta g_R}{A}   -  \frac{B(c_{2 \theta_W} \delta g_L    -  2 s_{\theta_W}^2  \delta g_R )}{A^2}    \,,
\end{align}
with $B =   - 1/2   + c_{2 \theta_W}$ and $A$ defined below \eqref{propeq}.

The semileptonic operators \eqref{o910} can in principle accommodate the anomalies observed in $b \to s$ transitions.   To analyze this, we reconstruct the likelihood for $b \to s \mu^+ \mu^-$ observables from the $1 \sigma$ and $2 \sigma$ contours in the $\C_9-\C_{10}$ plane provided in~\cite{Altmannshofer:2017fio}, assuming a bivariate normal distribution.    We obtain $(\C_{9},\C_{10}) = (-1.11,0.273)$ for the mean values, $\sigma_{C_9} = \sigma_{C_{10}} = 0.24$ for the standard deviation, and a correlation $\rho = 0.20$.    We also include in our analysis the ratios $R_K$ and $R_{K^*}$, using the general formulas derived in \cite{Celis:2017doq} and the experimental values reported in \cite{Aaij:2014ora,Aaij:2017vbb}.     Contributions to $b \to s \nu \bar \nu$ and $s \to d  \nu \bar \nu$ are related in this framework to those in $b \to s \mu^+ \mu^-$ due to the $\mathrm{SU(2)}_L$ gauge symmetry and the predictive flavour structure~\cite{Kamenik:2017tnu,Buras:2014fpa}.  Current bounds from $B$ and $K$ meson decays into final states with neutrinos do not set any relevant constraint in our framework.

\begin{figure}[htp]
\begin{center}{
\hspace{-10pt}\includegraphics[width=7.8cm]{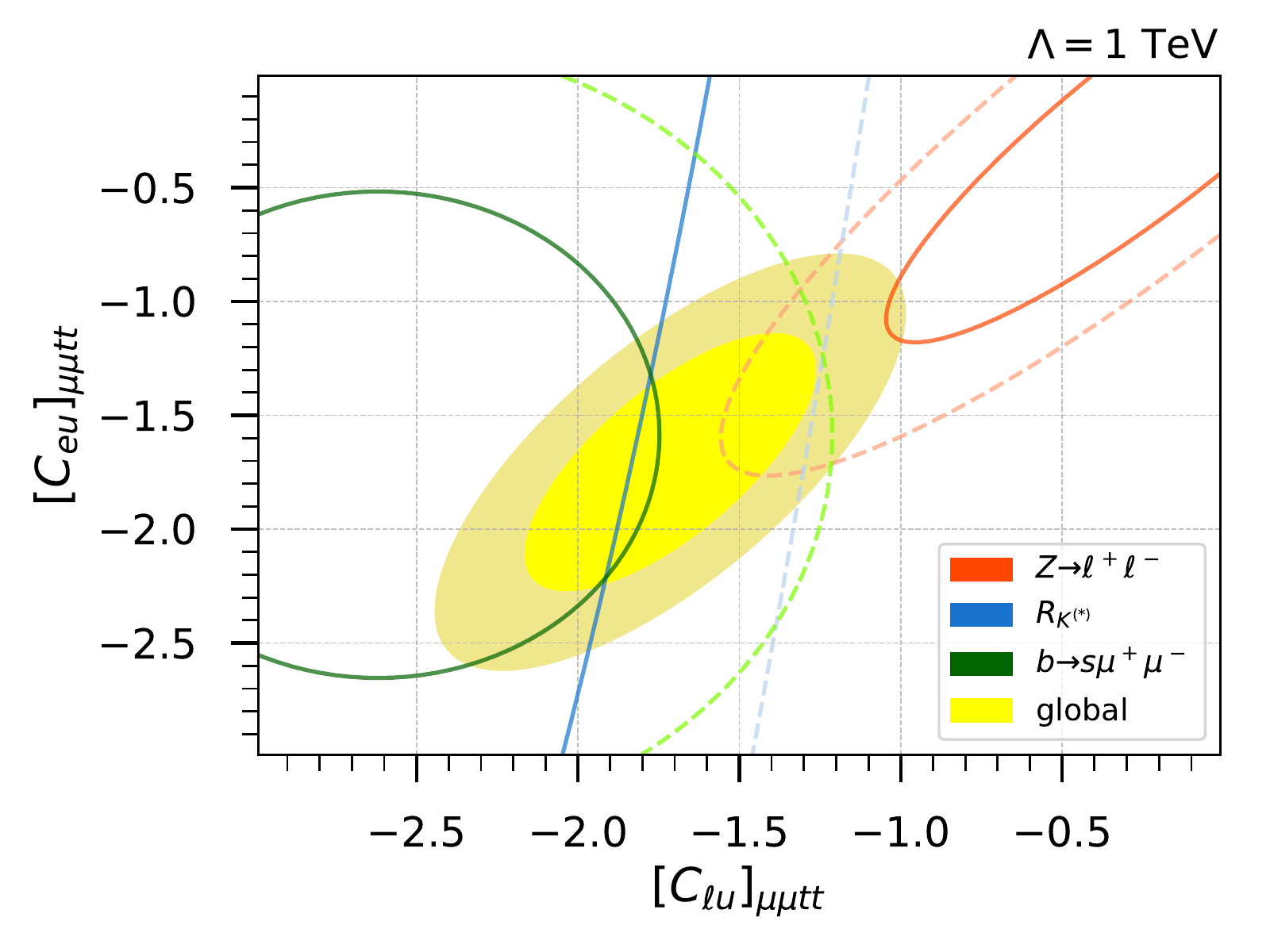}
~
\includegraphics[width=7.5cm]{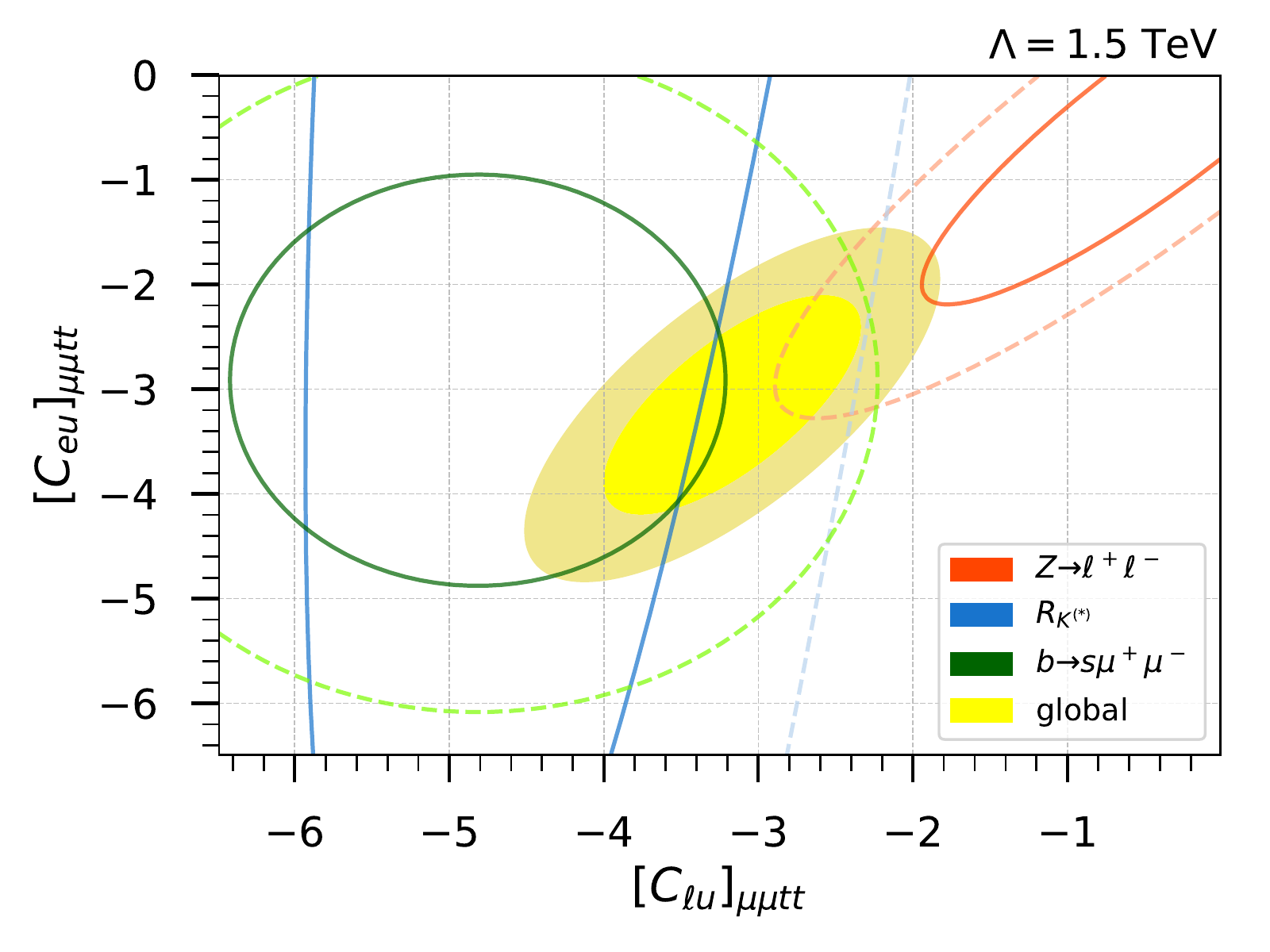}
~
\includegraphics[width=7.5cm]{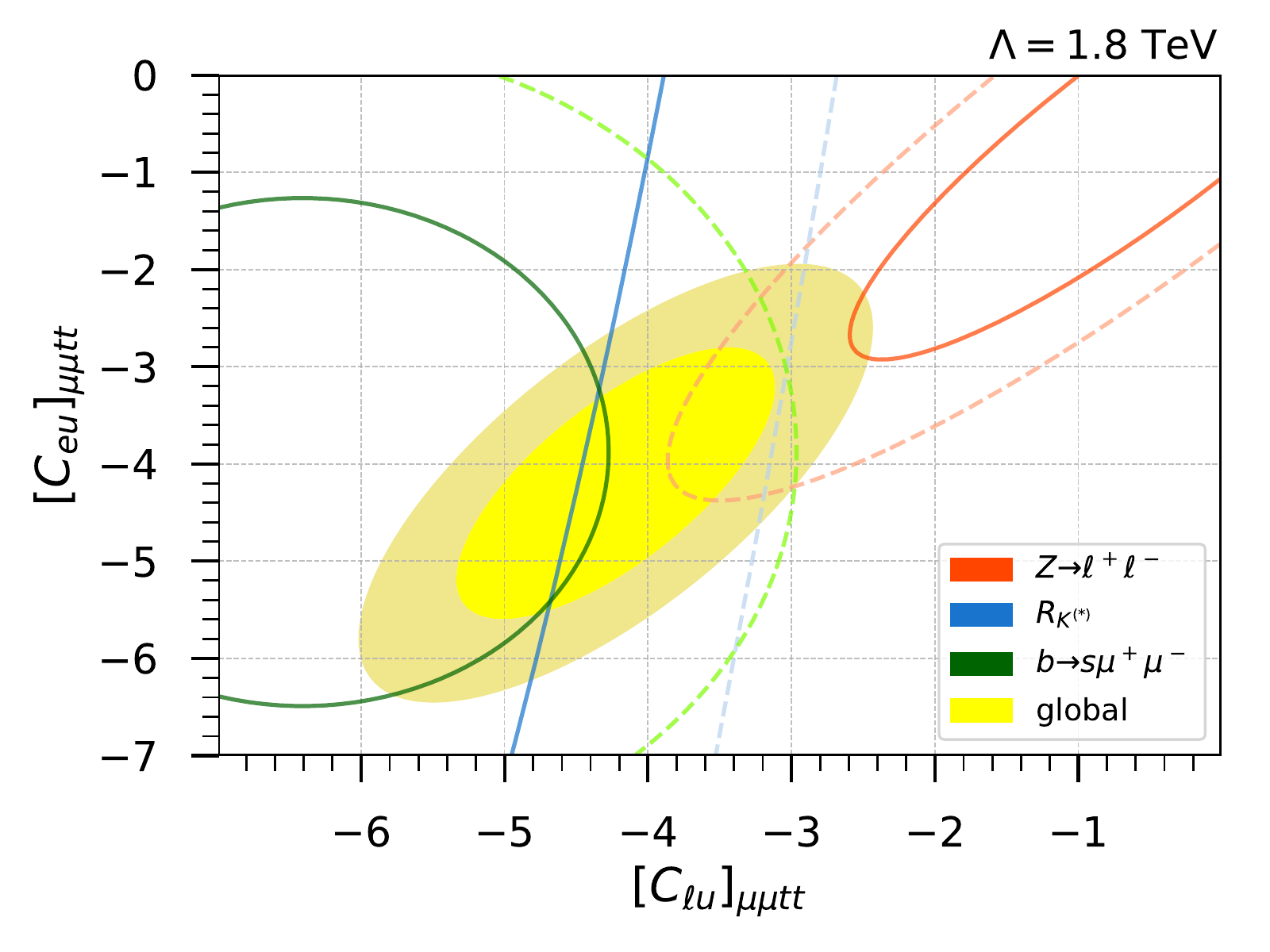}
\caption{\textit{Preferred regions at $68\%$ and $95\%$~confidence level (CL) in the $(\C_{\ell u}, \C_{eu})_{\mu \mu tt}$ plane from the global $\chi^2$ (yellow-filled), $b \to s \mu^+ \mu^-$ observables (green), $R_{K^{(*)}}$ (blue) and LEP-I measurements (red).   Three benchmark values of the high scale $\Lambda= 1,1.5,1.8$~TeV have been chosen.}  \label{figplots} } }
\end{center}
\end{figure}

A global $\chi^2$ function is built with all these observables.  The results of the fit are summarized in Table~\ref{tab:chi2} and in Figure~\ref{figplots}.  Table~\ref{tab:chi2}  shows the contributions to $\chi^2$ from each sector within the SM and at the minimum of the global $\chi^2$ for three benchmark values of $\Lambda$.    Figure~\ref{figplots}  shows the isocontours of $\Delta \chi^2   \equiv \chi^2 - \chi^2_{\rm{min}}  = \{2.3,5.99\}$ in the plane $\{\C_{\ell u}, \C_{eu}\}$ for the same benchmarks.          The preferred region by the global fit (shown in Figure~\ref{figplots} as a yellow ellipse) lies is the third quadrant along the direction $\C_{eu} \sim \C_{\ell u}$.    In this region, the NP contribution to the effective Hamiltonian for $b \to s \ell^+ \ell^-$ transitions enters mainly in the Wilson coefficient $\C_9$. 
  
One important observation is that the NP effects considered cancel accidentally for $\C_{eu} \sim \C_{\ell u}$ in the decay width for $Z \to \mu^+ \mu^-$, see Eq.~\eqref{propeq} ($c_{2 \theta_W} \simeq 2 s_{\theta_W}^2$). The leptonic asymmetry parameter ${\cal{A}}_{\mu}$ breaks this blind direction of the LEP-I $\chi^2$ to some degree, but a very strong correlation between these two variables remains.       We have compared the LEP-I  bounds we obtain with those derived using the results of~\cite{Efrati:2015eaa} and found good agreement.   For this comparison we use the following values reported in Ref.~\cite{Efrati:2015eaa}: $\delta g_L/2 = (0.1 \pm 1.1) \times10^{-3}$, $\delta g_R/2 = (0.0 \pm 1.3) \times10^{-3}$, with a correlation $\rho= 0.90$.    Another observation is that current data for $b \to s \ell^+ \ell^-$ and LEP-I show some slight tension within the framework analyzed here, which is reflected in the contribution of LEP observables to the $\chi^2$ in Table~\ref{tab:chi2}. The combined fit would be better if the deviations from the SM observed in $b \to s \ell^+ \ell^-$ transitions decrease slightly with future measurements.

\begin{table}[ht]
\caption{\textit{Contribution to the $\chi^2$ from each sector at the minimum of the global $\chi^2$ and in the SM. } \label{tab:chi2}  }
\begin{center}
\begin{tabular}{|c|c|c|c|c|}
    $\chi^2$  &     $b \to s \mu^+ \mu^-$  &  $R_{K^{(*)}}$   &   $Z \to \ell^+ \ell^-$  \\    \hline   
   ${\rm{SM}}$  &      25.8   &     22.5    &   0.5           \\ 
   $\Lambda = 1$~TeV  &     2.5 &   5  &   7.9  \\
      $\Lambda = 1.5$~TeV  &    2.5       &    5  &    7.8  \\
               $\Lambda = 1.8$~TeV  &   2.4             &    5    &    7.8  \\
\end{tabular}
\end{center}
\end{table}%

\section{Mediators}   \label{secmodels}

\begin{table}[b]
\caption{\textit{Possible mediators generating at tree-level the two relevant operators.   The $Z^\prime$ represents a vector boson in the singlet representation of the SM gauge group while the nomenclature used for the LQs corresponds to that in~\cite{Dorsner:2016wpm}.   The last row shows those for which the Wilson coefficients are negative, as required by the low-energy fit.} \label{tab:listmed}  }
\begin{center}
\begin{tabular}{c |c c c c c |}
 \ \          &\ \ $Z^\prime$\ \ &\ \ $S_1$\ \ &\ \ $R_2$\ \ &\ \ $\widetilde{U}_1$\ \ &\ \ $\widetilde{V}_2$\ \ \\\hline   
$[\mathcal{O}_{\ell u}]_{\mu \mu tt}$\ \ \ & \color{blue}\cmark & \color{sexyred}\xmark & \color{blue}\cmark & \color{sexyred}\xmark & \color{blue}\cmark \\
$[\mathcal{O}_{eu}]_{\mu \mu tt}$\ \  \   & \color{blue}\cmark & \color{blue}\cmark & \color{sexyred}\xmark & \color{blue}\cmark & \color{sexyred}\xmark \\
$\C_{\ell u},\,\C_{eu}<0$ \  & \color{blue}\cmark & \color{sexyred}\xmark & \color{blue}\cmark & \color{blue}\cmark  & \color{sexyred}\xmark \\
\end{tabular}
\end{center}
\end{table}%

 Different mediators can in principle generate the operators considered in Eq.~\eqref{twoop}. Taking into account the different irreducible representations of the Lorentz and SM gauge symmetry groups, one finds that there are only five different states that can generate these operators at tree-level~\cite{deBlas:2017xtg}, shown in Table~\ref{tab:listmed}.  The required size of the Wilson coefficients as well as their sign in Figure~\ref{figplots} provides important information about the possible models that can accommodate the anomalies, ruling out two of the possible mediators.
\subsection{$Z^{\prime}$ boson}  \label{seczprime}

One candidate mediator is a vector boson in the irreducible representation $Z_{\mu}^{\prime} \sim (1,1,0)$ of the SM gauge group $\mathrm{SU(3)}_C \times \mathrm{SU(2)}_L \times \mathrm{U(1)}_Y$.  Such state can arise in scenarios of strong dynamics behind EW symmetry breaking or in weakly coupled extensions of the SM with an extended gauge group~\cite{Langacker:2008yv}.   A model with a $Z^{\prime}$ boson coupling predominantly to right-handed top-quarks and to muons in order to explain the $b\to s \ell^+ \ell^-$ anomalies was presented in~\cite{Kamenik:2017tnu} and also analyzed in~\cite{Fox:2018ldq}.    We are interested in an interaction Lagrangian of the form
\begin{align}  \label{eq:zprime}
\mathcal{L} =   Z_{\alpha}^{\prime}  \left[  \bar \mu    \gamma^{\alpha}  (   \epsilon_L^{\mu \mu}   P_L   +  \epsilon_R^{\mu \mu}  P_R )   \mu  +      \epsilon_{R}^{tt}  \, \bar t  \gamma^{\alpha}  P_R   t     \right] \,.
\end{align}
Though it is not written here explicitly, a $Z^{\prime}$ interaction to muon neutrinos would also arise in common UV theories, as together with muons they make doublets of the $\mathrm{SU(2)}_L$ gauge symmetry of the SM. Integrating out the $Z^{\prime}$ boson from the theory at tree level gives rise to the matching conditions
\begin{align}
\C_{\ell u} &= -     \epsilon_{R}^{tt}      \epsilon_L^{\mu \mu}   \,, \qquad  \C_{eu} = -      \epsilon_{R}^{tt}      \epsilon_R^{\mu \mu}  \,.
\end{align}
The fact that the preferred region in Figure~\ref{figplots} lies around the line $\C_{eu}  \sim \C_{\ell u}$ implies $\epsilon_R^{\mu \mu} \sim      \epsilon_L^{\mu \mu}$.  Of the possible mediators in Table~\ref{tab:listmed}, the $Z^\prime$ boson is the only single state capable of simultaneously generating both operators in Eq.~\eqref{twoop} with the correct negative sign.

\subsection{Leptoquarks}  \label{seclqs}

Leptoquarks (LQ) are exotic colored particles mediating quark-lepton transitions~\cite{Buchmuller:1986zs,Dorsner:2016wpm}.  Such particles are known to arise in unification scenarios~\cite{Georgi:1974sy,Pati:1974yy} or in scenarios of strong dynamics behind the EWSB~\cite{Gripaios:2009dq}.  In contrast to the $Z^\prime$ case, no single LQ mediator can simultaneously generate both operators in Eq.~\eqref{twoop}.  Nonetheless, we find that a combination of a scalar and a vector LQ can reproduce the preferred region in Figure~\ref{figplots}. 

The two operators considered in \eqref{twoop} receive tree-level matching contributions by integrating out scalar and vector LQs with SM quantum numbers
\begin{align}
R_2 \sim (3,2,7/6) \qquad  \widetilde U_{1 \alpha} \sim (3,1,5/3) \,.
\end{align}

The interactions of these LQs with the fermions are described by the Lagrangian
\begin{align} \label{lqlag}
\mathcal{L}=&\;     \kappa_S \,  \bar t_{R}  \,     R_2^{T}  i \tau_2  \ell_{L \mu}     +  \kappa_{V} (  \bar t_{R} \gamma^{\alpha}  \mu_{R})  \widetilde U_{1 \alpha}   +\text{h.c.} \,.
\end{align}  
Here $\ell_{ L \mu} = (\nu_{\mu}, \mu_{L})^T$ is the lepton doublet, $\tau_2$ is the Pauli matrix and we have not written the interaction term $\bar q_L R_2 e_R$, which is allowed by the SM gauge symmetry. We will assume this term is forbidden by some underlying symmetry of the model for simplicity.   Integrating out the LQs at tree-level gives the matching conditions~\cite{delAguila:2010mx,Alonso:2015sja,Bobeth:2017ecx}
\begin{align} \label{myeqsv}
[\C_{\ell u}]_{\mu \mu tt}   &=  -  \frac{ |\kappa_S|^2 }{2}  \,, \qquad \, [\C_{e u}]_{\mu \mu tt} =   -   |\kappa_V|^2 \,.
\end{align}
If the LQs have similar mass, the preferred region in Figure~\ref{figplots} implies that $|\kappa_S| \sim \sqrt{2} |\kappa_V|$. Notice that the Wilson coefficients in \eqref{myeqsv} have the negative sign necessary to accommodate the low-energy fit, and that no other operators besides the ones in Eq.~\eqref{twoop} are generated at tree-level from integrating out these LQs.  The remaining two LQs in Table~\ref{tab:listmed} generating these Wilson coefficients~\cite{delAguila:2010mx,Alonso:2015sja,Bobeth:2017ecx}, $S_1 \sim (3^*,1,1/3)$ and  $\widetilde V_{2 \alpha} \sim (3^*,2,-1/6)$, lead to Wilson coefficients with the opposite signs as those in Eq.~\eqref{myeqsv} and are therefore not able to fit the experimental data.

Note that introducing a massive vector LQ with an explicit mass term spoils the renormalizability of the theory, contrary to a scalar LQ.  Introducing an ultraviolet origin for the vector LQ (for example from an spontaneously broken gauge theory) is necessary to calculate one-loop finite corrections to the matching at the high energy scale.

\section{High-$p_T$ phenomenology} \label{sechighpt}

\subsection{Limits on the $Z^\prime$ model}

We now turn to the phenomenological implications of the $Z^\prime$ mediator discussed in Sec.~\ref{seczprime}, assuming it has a mass around the TeV scale and vectorial coupling to muons $\epsilon_V^{\mu\mu}\equiv\epsilon_L^{\mu\mu}=\epsilon_R^{\mu\mu}$.  When extracting limits from the LHC we will focus on tree-level $Z^\prime$ exchanges and omit from our analysis loop-induced processes such as $gg \to g Z^{\prime}$. The latter are sensitive to details of the ultraviolet completion such as effects from heavy fermionic top-partners. These exotic fermions are not uncommon when trying to build an ultraviolet completion for the $Z^\prime$ model at hand and, while being too heavy to be directly produced on-shell they still may give non-negligible contributions to the production of the lighter $Z^\prime$ through loop-level non-decoupling effects, see Ref.~\cite{Kamenik:2017tnu,Fox:2018ldq} for more details. 
\begin{figure}[t]
\begin{center}{
\includegraphics[width=7.5cm]{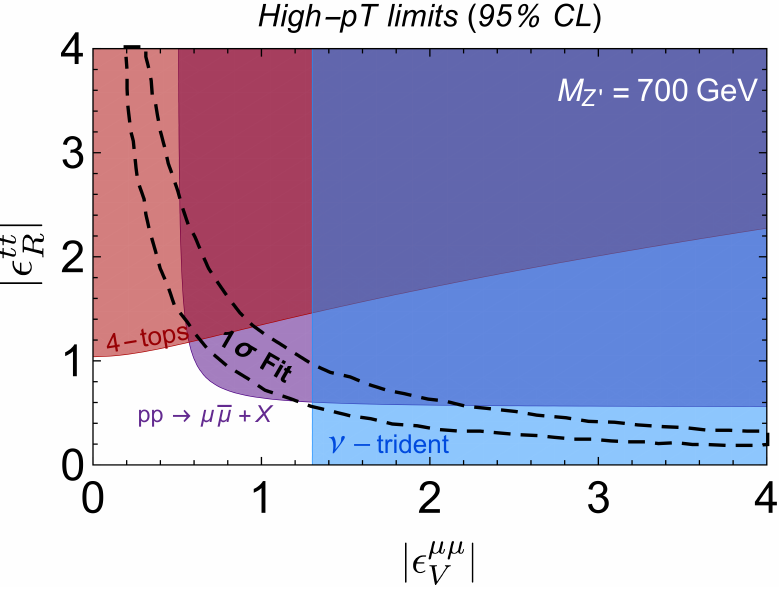}\\ \vspace{0.3cm}
\includegraphics[width=7.5cm]{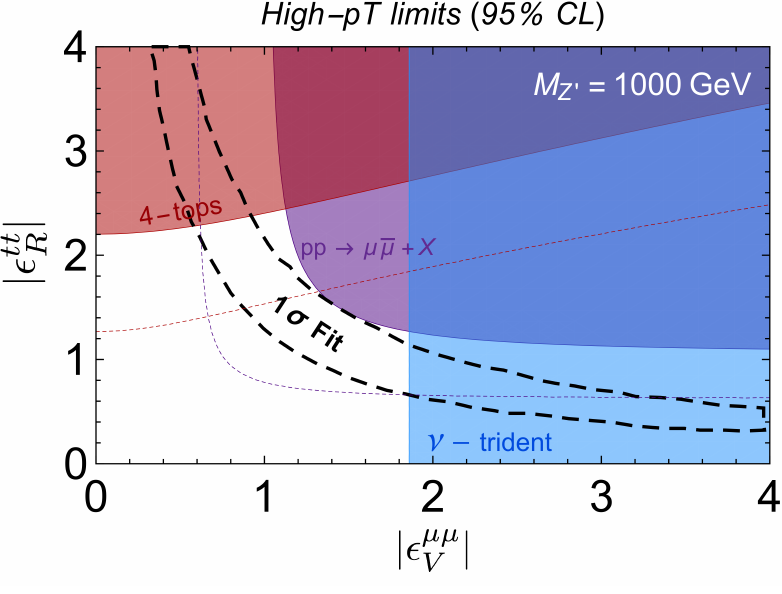}\\
\caption{\label{fig:ZprimeLHC} \textit{Summary of high-$p_T$ bounds for the $Z^\prime$ model. The red, purple and blue 95\% CL exclusion regions correspond to the LHC 4-top search, LHC di-muon tail search and the CCFR neutrino trident experiment, respectively. Dotted colored contours represent LHC bounds at a future luminosity of 300\,\invfb. The black dashed region corresponds to the $1\,\sigma$ global fit to $R_{K^{(*)}}$, LEP-I data and the $b\to s\mu\mu$ observables.} } }
\end{center}
\end{figure}

At tree level, the most important constraints come from $Z^\prime$ production in association with $t\bar t$ at the LHC. Once produced, the $Z^{\prime}$ boson can decay into muons, muon-neutrinos, and top-quarks. After neglecting small lepton masses, the partial decay widths for these channels are given by
\begin{align}\label{eq:zwidths}
&\Gamma(Z^{\prime}\to \mu\bar\mu)  \simeq \frac{   M_{Z^{\prime}}  }{      24 \pi }   (   |\epsilon_L^{\mu \mu}|^2 +  |\epsilon_R^{\mu \mu}|^2 )  \,, \nonumber \\
&\Gamma(Z^{\prime}\to t \bar t)  \simeq \frac{   \lambda^{1/2}(1,z_t,z_t)  N_C  M_{Z^{\prime}}  }{      24 \pi }  (1-z_t)      |\epsilon_R^{tt}|^2  \,, \nonumber \\ 
&\Gamma(Z^{\prime}\to \nu_{\mu} \bar \nu_{\mu})  \simeq \frac{    M_{Z^{\prime}}  }{      24 \pi }     |\epsilon_L^{\mu \mu}|^2   \, ,
\end{align}
where $z_t = m_t^2/M_{Z^{\prime}}^2$, $N_C=3$, and $\lambda$ represents the K\"all\'en function. Each of these decay channels give rise to three complementary LHC signatures: $t\bar t\mu\bar\mu$, $t\bar tt\bar t$ and $t\bar t+\slashed{E}_T$. In order to set limits in the coupling plane $\{\epsilon_V^{\mu\mu},\epsilon_R^{tt}\}$ of the model we have recasted a set of existing LHC searches for two benchmark masses of $M_{Z^\prime}=0.7$~TeV and $M_{Z^\prime}=1$~TeV. For each benchmark mass, the $1\,\sigma$ favored region fitting the $b\to s \ell^+\ell^-$ anomalies and the $\mbox{LEP-I}$ measurements is given by the black dashed contours in Figure~\ref{fig:ZprimeLHC}. Limits for the process $pp\to t\bar t Z^\prime\to t\bar t\mu\bar\mu$ were extracted from the generic $Z^\prime$ di-muon resonance search by ATLAS~\cite{Aaboud:2017buh} (Sec.10.3) at 36.1\,\invfb, assuming a detector acceptance of $40\%$ and a decay width dominated by the three channels in Eq.~\eqref{eq:zwidths}. The 95\% CL exclusion limits from this search are shown in the purple region in Figure~\ref{fig:ZprimeLHC}. Projections to a higher luminosity of 300\,\invfb are also given by the dotted purple contour in the same figure. For the process $pp\to t\bar t Z^\prime\to t\bar tt\bar t$ we used the current best upper limit on the SM four-top cross-section by CMS~\cite{Sirunyan:2017roi} at 35.9\,\invfb of data. The $2\sigma$ exclusion bound is given by the red region in Figure~\ref{fig:ZprimeLHC}. Projections to 300\,\invfb, given by the dotted red contour, were estimated using the multi-leptonic analysis performed in Ref.~\cite{Alvarez:2016nrz}, where the 95\% CL upper limit on the SM cross-section was found to be approximately $\sigma^{\text{SM}}_{t\bar t t\bar t}<23$~fb. Notice that a dedicated resonance search for this channel can considerably improve this bound (especially at higher luminosities) if a high-mass cut is applied on the top-quark decay products or a top-tagger is used in order to improve sensitivity to the boosted tops from the decaying resonance, see also Ref.~\cite{Kim:2016plm}.       

Another relevant probe of the $Z^{\prime}$ boson is the neutrino trident production~\cite{Altmannshofer:2014pba}.    The process $\nu_\mu \gamma^*\to\nu_\mu\mu\bar\mu$ occurring in a fixed target from a highly energetic neutrino beam gives important limits on the $Z^\prime$ boson coupling to muonic currents for a wide range of $Z^\prime$ masses.  These constraints will be complementary to those from the LHC. The cross-section for this process normalized by the SM prediction is given by~\cite{Altmannshofer:2014pba}
\begin{align}
\dfrac{\sigma^{\text{NP}}_{\nu_\mu\mu\bar\mu}}{\sigma^{\text{SM}}_{\nu_\mu\mu\bar\mu}} = \dfrac{1+\left(1+4 s_{\theta_W}^2 +\dfrac{2v^2\,(\epsilon_V^{\mu\mu})^2}{M_{Z^\prime}^2}\right)^{2}}{1+(1+4 s_{\theta_W}^2)^2}.
\end{align}
This quantity has been measured at CCFR to be $\sigma^{\text{NP}}_{\nu_\mu\mu\bar\mu}/\sigma^{\text{SM}}_{\nu_\mu\mu\bar\mu}=0.82\pm0.28$~\cite{Mishra:1991bv}, giving a strong constraint on the $Z^\prime$ muonic couplings. The $2\,\sigma$ upper limit on $\epsilon_V^{\mu\mu}$ is represented by the vertical blue region in Figure~\ref{fig:ZprimeLHC}.   

We observe that all the constraints are complementary and exclude different regions of the available parameter space in Figure~\ref{fig:ZprimeLHC}.      For $M_{Z^{\prime}}=0.7$~TeV, the preferred $1\sigma$ region from the global fit of flavour and LEP observables is already excluded, with each of the different constraints considered playing an important role.   For $M_{Z^{\prime}} = 1$~TeV, an allowed region remains centered around the point $\{|\epsilon_{V}^{\mu \mu}|,|\epsilon_R^{tt}|\} = \{1.2, 1.2\}$.  Future searches from the LHC with $300$~fb$^{-1}$ will be sensitive to this region.

\subsection{ Limits on the $\mathbf{R_2}$ plus $\mathbf{\widetilde{U}_1}$ model}\label{sec:R2U1highpt}

\begin{figure}[h]
\begin{center}{
\includegraphics[width=6cm]{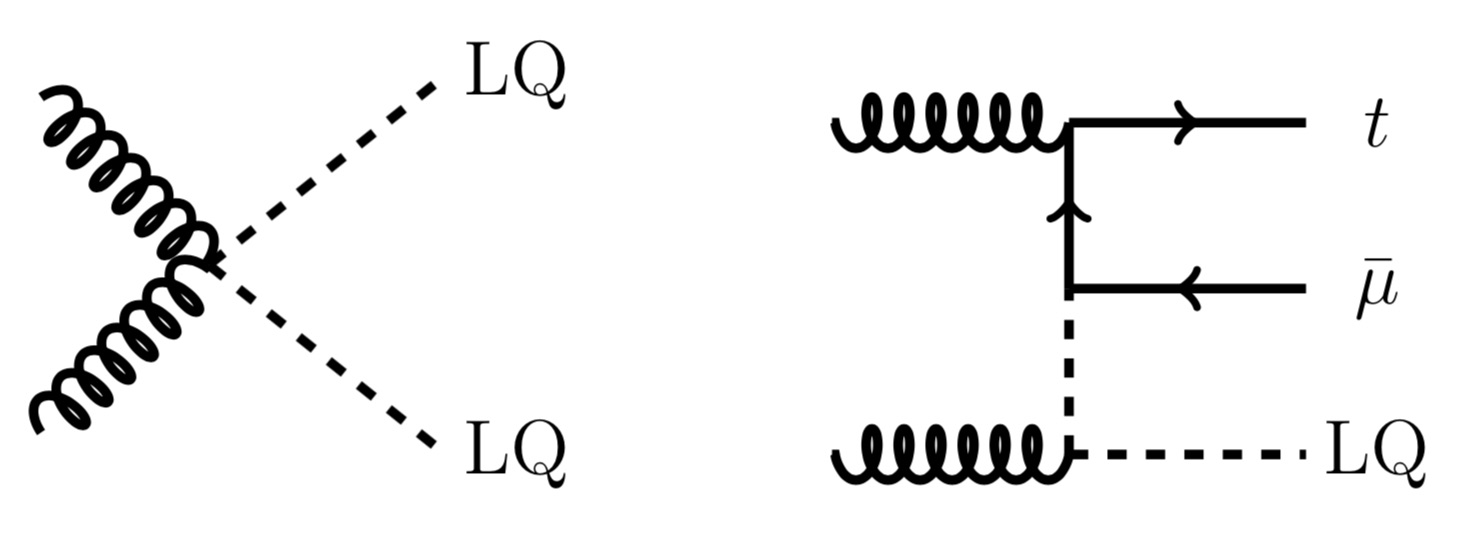}
\caption{\textit{Representative diagrams for the QCD LQ pair production (left) and for the single LQ production mode $gg\to \widetilde{U}_1(R_2)\,t\mu$ (right). }\label{feynlq}} }
\end{center}
\end{figure}

For this model the most important LHC bound comes from LQ pair production $gg\, (q\bar q)\to \widetilde{U}_1^\dagger \widetilde{U}_1,\,R_2^\dagger R_2$. The consequences of having the interactions in Eq.~\eqref{lqlag} plus a negligible top-quark PDF for the proton are: (i) LQ pair production is independent of the size of the couplings $\kappa_{S,V}$, hence driven completely by QCD interactions (see Figure~\ref{feynlq} (left) for a representative Feynman diagram), (ii) the absence of all $2\to2$ single LQ production channels of the form $qg\to\mathrm{LQ}\, \ell$ at the LHC and (iii) the absence of $q\bar q\to\ell\bar\ell$ production via LQ exchange in the $t$-channel. The only relevant process at the LHC at leading order besides QCD pair production is the $2\to3$ single LQ production mode $gg\to \widetilde{U}_1(R_2)\,t\mu$ shown in Figure~\ref{feynlq} (right). This last process only becomes competitive with LQ pair production if the couplings $|\kappa_{S,V} |\gsim1$ are large enough to overcome the $2\to3$ body phase space suppression.

The scalar leptoquark doublet $R_2$ (see Sec.~\ref{seclqs}) when decomposed into its $\mathrm{SU(2)}_L$ components $R_2^T = (R^{5/3},R^{2/3})$, where the upper indices denote the electric charge ($Q = Y + T_3$), gives the following interactions
\begin{align} \label{myEq}
\mathcal{L} \supset&\             \kappa_S \left[  \,  \bar t_{R}  \,  \nu_{\mu}     R^{2/3}    -    \,  \bar t_{R}  \,    \mu_{L} R^{5/3}   \right]  +\text{h.c.} 
\end{align}  
In this case the branching ratio for each state reads
\begin{align}
\beta(R^{2/3} \to t \nu_{\mu}) = 1 \,, \quad   \beta(R^{5/3} \to t \mu  ) = 1  \,,
\end{align}
where $\beta(X\!\to\! Y)\! \equiv\! \Gamma(X\! \to\! Y)/\Gamma_{\rm{tot}}$. The vector leptoquark singlet in Eq.~\eqref{lqlag} has a branching fraction of $\beta(\widetilde U_{1 \alpha}  \to t \mu) = 1$. 

We derive constraints on the LQs in the $t\bar t \mu\bar\mu$ channel by a recast of recent SUSY searches by ATLAS in the four-lepton~\cite{Aaboud:2018zeb} and same-sign di-lepton $+$ tri-lepton channels~\cite{Aaboud:2017dmy}. We also derive bounds by a recast of an inclusive di-muon resonance search~\cite{Aaboud:2017buh}. In order to estimate the number of signal events in each signal region, we first wrote UFO model files for $R_2$ and $\widetilde{U}^\alpha_1$ using {\tt FeynRules}~\cite{Alloul:2013bka} and generated large LQ pair production samples in {\tt MadGraph5}~\cite{Alwall:2014hca}. The decays of the tops into all channels where performed directly in {\tt Pythia8}~\cite{Sjostrand:2014zea} as well as parton showering and hadronization. Finally, for each search, detector effects where simulated with {\tt Delphes3}~\cite{deFavereau:2013fsa}. Selection cuts for the signal regions for each search were applied to the samples in order to extract the signal efficiencies.

For the scalar LQ pair production cross-section we used the NLO parametric representation given in \cite{Mandal:2015lca}. For the vector LQ, the calculation of the pair production cross-section requires some assumptions about the underlying theory generating such state. The vector LQ-gluon interactions are parametrized by the following terms in the Lagrangian~\cite{Blumlein:1996qp}
\begin{align}
- \frac{1}{2} \widetilde{U}_{\alpha\beta}^\dagger\widetilde{U}^{\alpha\beta} -i   g_s\left[ \omega_G   \widetilde{U}_{1\alpha}^\dagger G^{\alpha\beta} \widetilde{U}_{1\beta}    +  \frac{\lambda_G}{M_{\widetilde{U}}^2}   \widetilde{U}_{\sigma \mu}^\dagger  G^{\nu \sigma}   \widetilde{U}_{\nu}^{\,\mu}      \right]  \,,
\end{align}
where  $\widetilde{U}^{\alpha\beta}\equiv D^\alpha\widetilde{U}^\beta_1\ - D^\beta\widetilde{U}^\alpha_1$, $G^{\alpha \beta}$ represents the gluonic field strength tensor and $D_{\alpha}$ is the SM gauge covariant derivative.    The parameters $\omega_G$ and $\lambda_G$ depend on the nature of the vector LQ. In our analysis we will assume $\widetilde{U}^\alpha_1$ to be a fundamental gauge boson arising from an extended gauge group. This choice fixes $\omega_G=1$ and $\lambda_G=0$. For this benchmark, the production cross-section for the vector LQ was calculated with {\tt MadGraph5} at leading order in QCD. We cross-checked our results with Ref.~\cite{Belyaev:2005ew}. Note that the production cross-section for the vector LQ is a factor of $\sim\cO(10)$ larger with respect to that of a scalar LQ with the same mass

For the SUSY searches, we used the 95\% CL limits provided by ATLAS on the number of allowed NP events in each signal region. Of all the SUSY searches, we found that the signal region {\tt Rpc3L1bH} of the tri-lepton search \cite{Aaboud:2017dmy} gives the best SUSY limits on the LQ masses: $M_R \gtrsim 1180$~GeV and $M_{\widetilde U} \gtrsim1720$~GeV. 

\begin{figure}[htp]
\begin{center}{
\includegraphics[width=7.5cm]{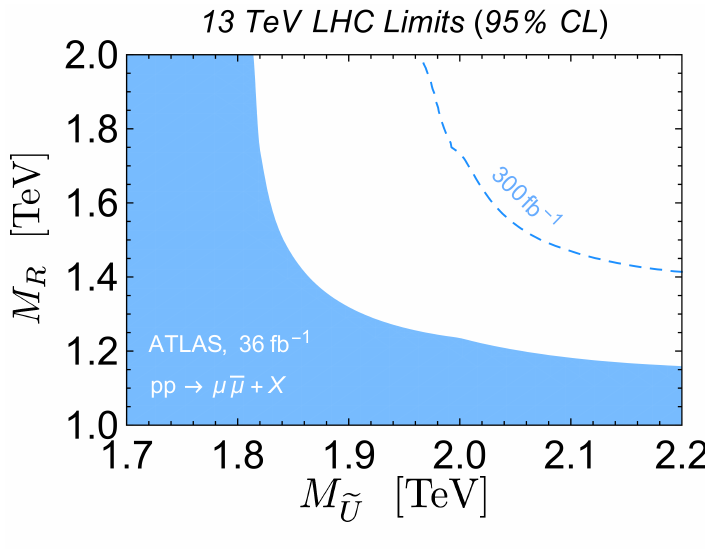}\\
\caption{\textit{Excluded mass region for the $R_2$ plus $\widetilde U_1$ model from the LHC di-muon tail search at $36.1$\,\invfb. The dashed contour shows the projected limit to 300\,\invfb of LHC data.} \label{fig:massLQs}     } }
\end{center}
\end{figure}

Finally, we turn to the inclusive di-muon tail search~\cite{Aaboud:2017buh}. The effect of the LQ resonant decay into $t\mu$ pairs is to modify the high-$p_T$ tails of the di-muon invariant mass spectrum. We compare signal and background events above an invariant mass cut of $m_{\mu\mu}>1200$~GeV (we find this value of the cut to be optimal for LQs above $1$~TeV) and perform a statistical analysis by log-likelihood minimization to extract the 95\% CL limits. In Figure~\ref{fig:massLQs} we show the excluded region in the $\{M_{\widetilde U},M_{R}\}$ plane from this search at 36.1\,\invfb \,  in red and a high luminosity projection with 300\,\invfb \, of data is given by the dashed red contour. Notice that having $M_{R} \sim M_{\widetilde U}$, for example, is allowed for masses above $1.9$~TeV. In appendix~\ref{appendixA} we give bounds on generic scalar and vector LQs decaying to $t\mu$ as a function of the branching fraction $\beta$ for one LQ at a time. These results from the SUSY tri-lepton search and the $pp\to \mu\bar\mu+X$ tails give the most stringent bounds up to date for this channel. For the pair production of the scalar LQ component $R^{2/3}$, we use a dedicated search by CMS~\cite{CMS:2018bhq} in the $t\bar t\nu\bar\nu$ channel at 35.9\,\invfb. This search however sets a weaker limit on the $R_2$ mass, $M_R>1020$~GeV.

\section{New physics in the electron channel}  \label{secenp}

\begin{figure}[ht]
\begin{center}{
\includegraphics[width=7.5cm]{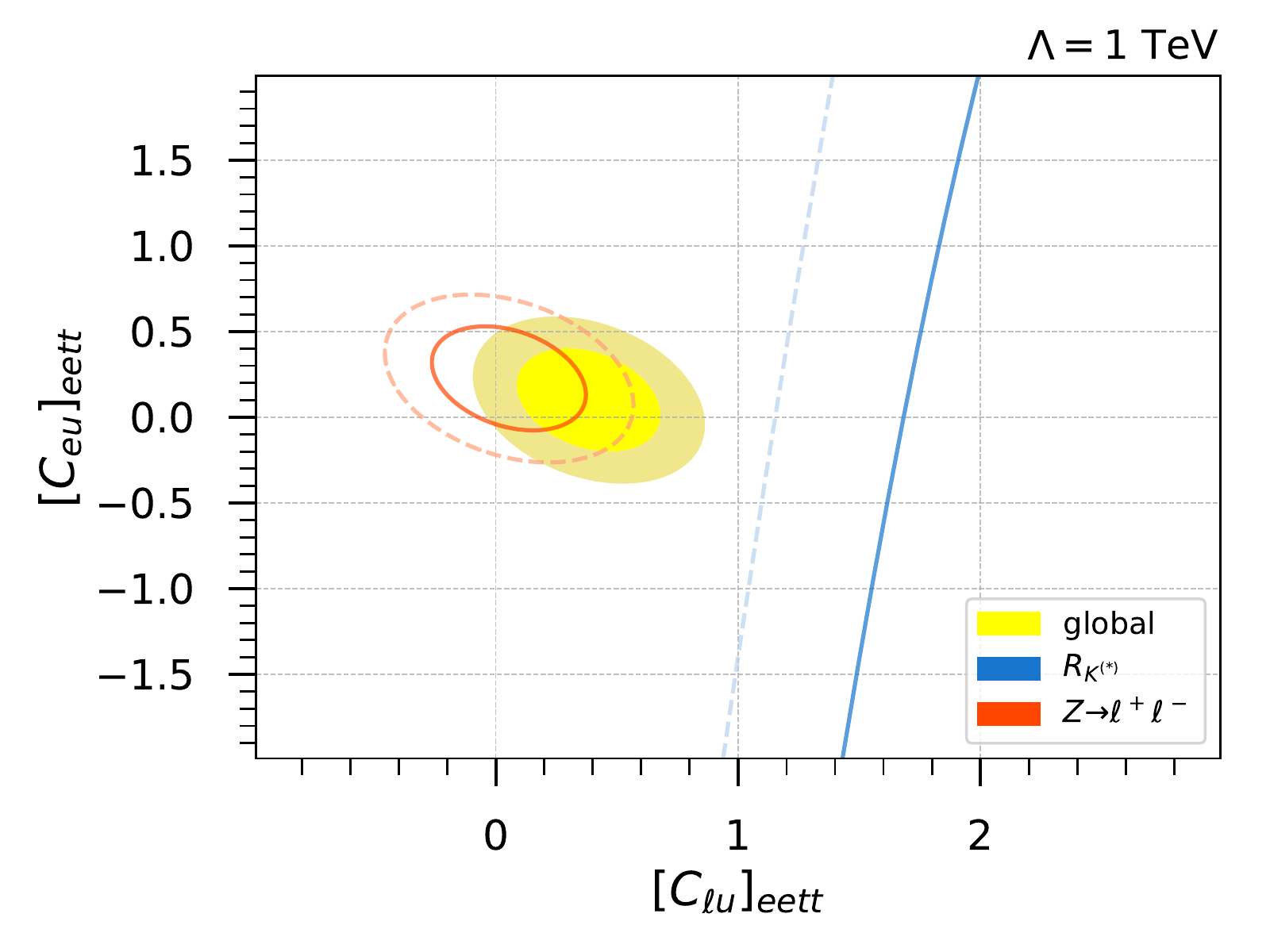}
\caption{\textit{Scenario with NP in the electron channel.  Preferred region at $68\%$ and $95\%$~CL in the $(\C_{\ell u}, \C_{eu})_{eett}$ plane from the global $\chi^2$ (yellow-filled), $R_{K^{(*)}}$ (blue) and LEP-I measurements (red).   The high scale $\Lambda$ has been fixed at $1~$TeV. }  \label{fig4}  } }
\end{center}
\end{figure}

A similar analysis can be performed assuming that NP in the operators \eqref{twoop} affects electrons instead of muons.  Obviously, in this case it is not possible to accommodate the anomalies in $b \to s \mu^+ \mu^-$, but one could still generate the required deviations in the ratios $R_{K^{(*)}}$.     In this case we use the measurement (see Table 7.4 in~\cite{ALEPH:2005ab})
\begin{align}
{\cal{A}}_{e} = 0.1514 \pm 0.0019 \,,
\end{align}
for the leptonic asymmetry parameter. The results of the global fit are summarized in Table~\ref{tab:chi2b} and Figure~\ref{fig4}. An important tension between the LEP-I bounds and the $R_{K^{(*)}}$ measurements is found due to more precise determination of the leptonic asymmetry parameter in this case.

\begin{table}[ht]
\caption{\textit{Scenario with NP in the electron channel. Values of $\chi^2$ within each sector at the global minimum and in the SM.}   \label{tab:chi2b} }
\begin{center}
\begin{tabular}{|c|c|c|c|}
    $\chi^2$  &    $R_{K^{(*)}}$   &   $Z \to \ell^+ \ell^-$  \\    \hline  
       SM  &       22.5  &    1.8 \\ 
   $\Lambda = 1$~TeV   &    16.6   &   2.7  \\
\end{tabular}
\end{center}
\end{table}
\section{Discussion of the results}  \label{secdisc}

$\bullet$  In Sec.~\ref{secmodels} we proposed two possible scenarios that generate the pattern of NP on which we are interested (a $Z^{\prime}$ boson or a combination of two leptoquarks). One can also consider different scenarios mixing these two.   For instance, a $Z^{\prime}$ with right-handed coupling to muons could be combined with the scalar leptoquark $R_2 \sim (3,2,7/6)$, in order to generate the two Wilson coefficients $\C_{\ell u}$ and $\C_{e u}$.  Alternatively, one could consider a $Z^{\prime}$ boson with left-handed coupling to muons combined with the vector leptoquark $\widetilde U_{1\alpha}$.

$\bullet$ Ref.~\cite{Celis:2017doq} performed a model independent analysis based on the SMEFT.   It was advocated that having $[\C_{\ell u}]_{\mu \mu tt} \sim - \mathcal{O}(1)$ for $\Lambda \sim 1$~TeV can provide a viable explanation of the $b \to s \ell^+ \ell^-$ anomalies.    In this work we have performed a similar analysis, including a more careful treatment of the LEP-I constraints.      We have included the required one-loop matching corrections at the EW scale that are relevant to estimate $Z \to \mu^+ \mu^-$ in this framework.     From our analysis, we find that $[\C_{\ell u}]_{\mu \mu tt} \sim - \mathcal{O}(1)$ for $\Lambda \sim 1$~TeV has some important tension with LEP-I measurements and a better solution is to have $[\C_{\ell u}]_{\mu \mu tt} \sim  [\C_{eu}]_{\mu \mu tt}  \sim  - \mathcal{O}(1)$ for $\Lambda \sim 1$~TeV.

\begin{figure}[t]
\begin{center}{
\includegraphics[width=7.5cm]{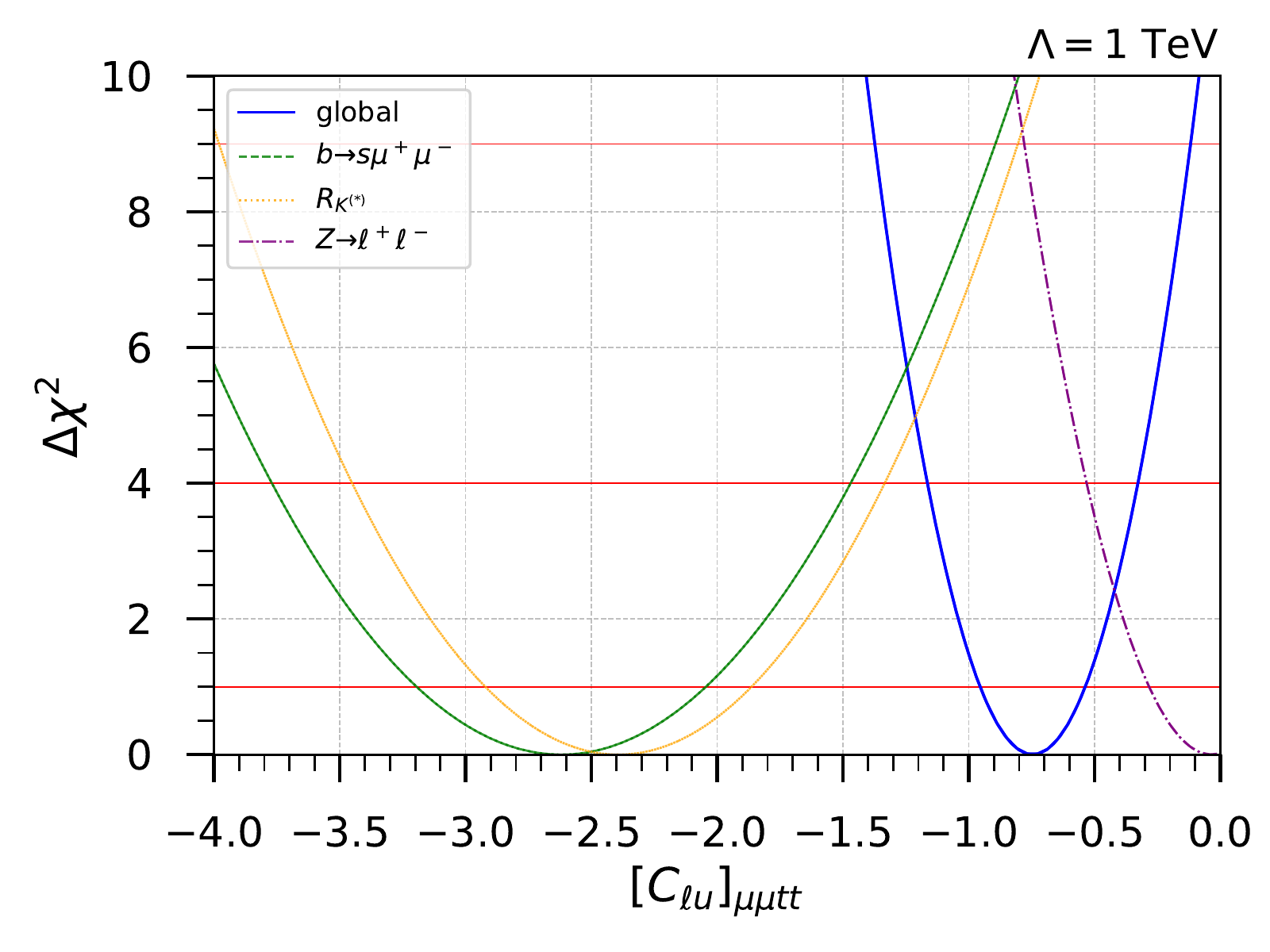}\\
\includegraphics[width=7.5cm]{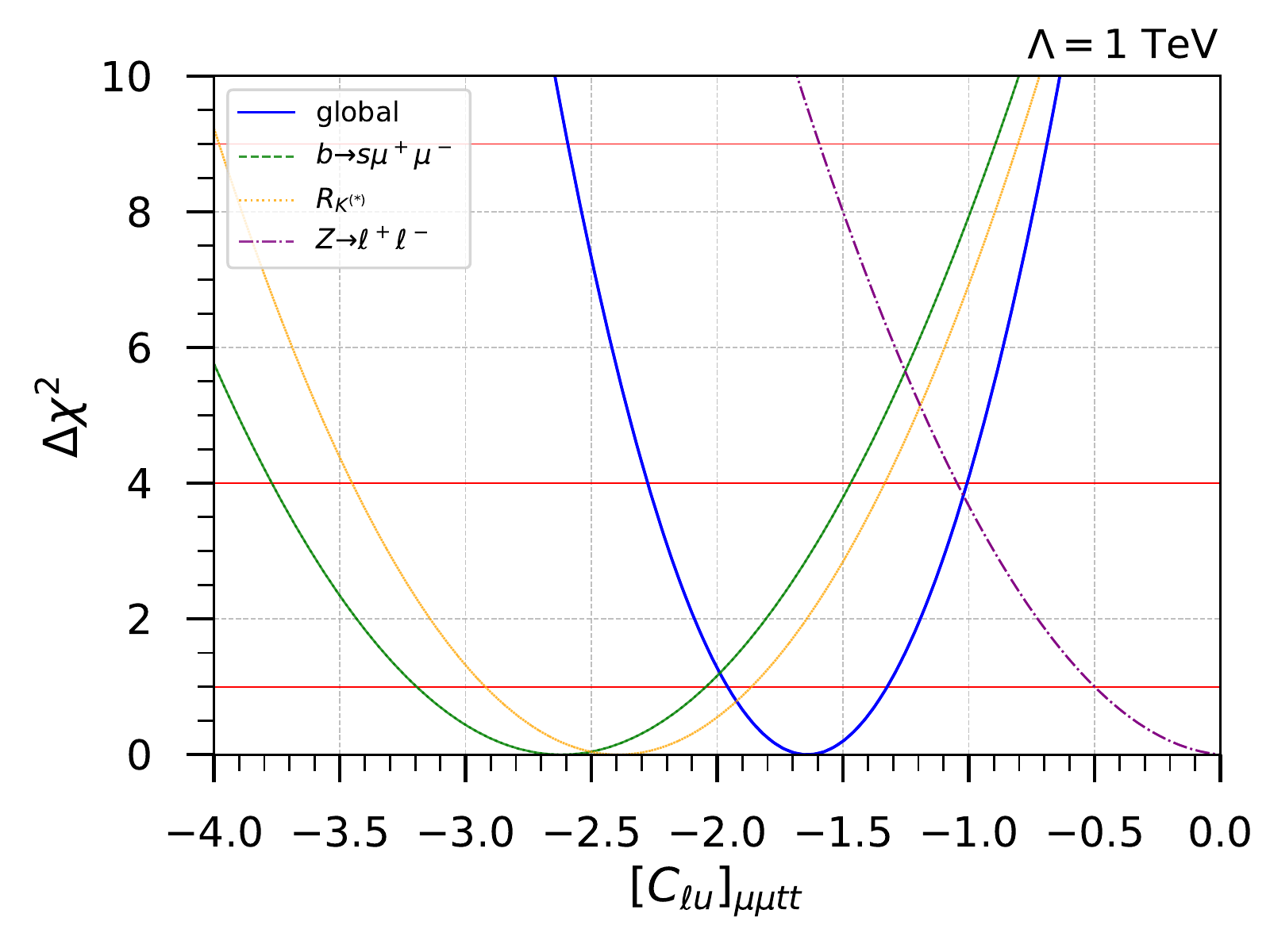}
\caption{\textit{Values of $\Delta \chi^2 = \chi^2 - \chi^2_{\rm{min}}$ against the Wilson coefficient $[\C_{\ell u}]_{\mu \mu tt}$, taking $[\C_{eu}]_{\mu \mu tt}=0$.   Horizontal lines show the values of $\Delta \chi^2=1,4,9$.   The high scale $\Lambda$ has been fixed to $1$~TeV.   Upper plot:     Results obtained in this work including the LEP-I measurements in \eqref{zdec} and \eqref{Amuexp}.   Lower plot:   Results obtained using the bounds derived in Ref.~\cite{Efrati:2015eaa} for $(\delta g_L,\delta g_R)$, but without taking into account the correlation. } \label{figplots2}     } }
\end{center}
\end{figure}

We have verified that the finite corrections to ${Z \to \mu^+ \mu^-}$ not included in Ref.~\cite{Celis:2017doq} are small and cannot explain the discrepancy. We find that the reason of the discrepancy was the large correlation ($\rho = 0.9$) between $(\delta g_L,\delta g_R)$ from Eq.~\eqref{Zlag}, which was not taken into account in~\cite{Celis:2017doq} when using the bounds from~\cite{Efrati:2015eaa}.   To illustrate this, we show in Figure~\ref{figplots2} the values of $\Delta \chi^2= \chi^2 - \chi^2_{\rm{min}}$ as a function of the Wilson coefficient $[\C_{\ell u}]_{\mu \mu tt}$ assuming $[\C_{e u}]_{\mu \mu tt}=0$.   Each sector included in the fit as well as the global $\chi^2$ are shown.         On the upper plot we show our results including the LEP-I measurements in \eqref{zdec} and \eqref{Amuexp}.   In the lower plot we show what happens when one uses instead the bounds from Ref.~\cite{Efrati:2015eaa} for $(\delta g_L,\delta g_R)$, without taking into account the correlation.    In Table~\ref{tab:chi2comp} we show the values of $[\C_{\ell u}]_{\mu \mu tt}$ at the minimum of the $\chi^2$ for each sector, taking $\Lambda = 1$~TeV.  When using the results from Ref.~\cite{Efrati:2015eaa}, missing the correlation between $(\delta g_L,\delta g_R)$ has the effect of reducing considerably the tension between LEP-I and $b \to s \ell^+ \ell^-$.     As remarked in Sec.~\ref{seclowph}, using the bounds from \eqref{zdec} and \eqref{Amuexp} leads to very similar results to taking the bounds on $(\delta g_L,\delta g_R)$ derived in Ref.~\cite{Efrati:2015eaa} if the correlation is included.

\begin{table}[ht]
\caption{\textit{Values of  $[\C_{\ell u}]_{\mu \mu tt}$ at the minimum of the $\chi^2$ for each sector fixing $\Lambda = 1$~TeV.} \label{tab:chi2comp}  }
\begin{center}
\begin{tabular}{|cc|c|c|c|c|} \hline 
       &  &     $b \to s \mu^+ \mu^-$  &  $R_{K^{(*)}}$   &   $Z \to \ell^+ \ell^-$  & global  \\    \hline   
  this work &$\C_{\ell u}$  &      -2.6     &   -2.4   &  -0.03   &  -0.75    \\
             &  $\chi^2$ &        5      &      2.1  &   0.5    &36  \\ \hline
           Ref.~\cite{Celis:2017doq}          &$\C_{\ell u}$  &      -2.6   &   -2.4   &  0   &  -1.6    \\
             &  $\chi^2$ &          5   &      2.1  &   0   &    21.6 \\ \hline
\end{tabular}
\end{center}
\end{table}%

$\bullet$ By integrating out the $Z^{\prime}$ of Eq.~\eqref{eq:zprime} at tree level one generates matching contributions to four-lepton and four-top operators.    The four-top operator $(\bar t_R \gamma_{\mu}   t_R )  (\bar t_R  \gamma^{\mu}  t_R )$ mixes at the two-loop level with the $\Delta F =2$ operator $(\bar b  \gamma_{\mu}  P_L  s  )(\bar b  \gamma_{\mu}  P_L  s )$, giving rise to $B_{s,d}$ meson mixing.  This new physics contribution to the $B_{s,d}$ meson mixing amplitude has the same CKM suppression as the SM contribution so that no new CP-violating phases are introduced.     For a $Z^{\prime}$ boson around $1$~TeV we obtain the bound $|\epsilon_{R}^{tt}|\lesssim 11$ from the measured mass differences $\Delta M_{s,d}$~\cite{Bazavov:2016nty}.  This  bound is much weaker than the one derived from $pp\to t\bar t Z^\prime\to t\bar tt\bar t$ at the LHC.  The situation is very different compared to the $Z^{\prime}$ models with tree level flavour violating couplings, for which stringent limits are derived from $B$ mixing, see for instance~\cite{DiLuzio:2017fdq,Alda:2018mfy}.

$\bullet$ It is important to stress that using the results derived in Sec.~\ref{seclowph} within the EFT framework to infer possible ultraviolet completions should be done carefully.     We can illustrate the possible subtleties with an extension of the SM with a $\mathrm{U(1)}^{\prime}$ gauge symmetry.    After spontaneous symmetry breaking of the $\mathrm{U(1)}^{\prime}$ symmetry, the $Z^{\prime}$ dynamics is described by the Lagrangian
\begin{align}
\mathcal{L} = - \frac{1}{4}  Z^{\prime \mu \nu}  Z^{\prime}_{\mu \nu}   + \frac{1}{2}  M_{Z^{\prime}}^2  Z^{\prime \mu}  Z_{\mu}^{\prime}   -  \frac{\kappa}{2}  B^{\mu \nu}  Z^{\prime}_{\mu \nu}  +   g_{Z^{\prime}}   Z^{\prime}_{\mu}  J_f^{\mu} \,,
\end{align}
with $Z^{\prime}_{\mu \nu}  = \partial_{\mu}  Z^{\prime}_{\nu}  -  \partial_{\nu}  Z^{\prime}_{\mu}$, $g_{Z^{\prime}}$ the $\mathrm{U(1)}^{\prime}$ gauge coupling, and $J_f^{\mu}$ representing the associated fermion current.  The term proportional to $\kappa$ is the kinetic mixing between the two abelian factors of the gauge group.       Integrating out the $Z^{\prime}$ field at tree-level gives rise to the dimension six effective Lagrangian (see for instance \cite{Henning:2014wua})
\begin{align}
\mathcal{L}_{\rm{eff}}  =  \frac{ g_{Z^{\prime}}  \kappa  }{M_{Z^{\prime}}^2}  (\partial_{\nu}   B^{\mu \nu}  )   J_{f \mu}    -  \frac{ g_{Z^{\prime}}^2}{2  M_{Z^{\prime}}^2}  (J_{f}^{\mu})^2   - \frac{\kappa^2}{2 M_{Z^{\prime}}^2}  (  \partial_{\nu}   B_{\mu \nu} )^2   \,,
\end{align}
which can be brought to the Warsaw operator basis using the SM equations of motion~\cite{Grzadkowski:2010es}.    After doing this, one obtains matching contributions to the operators $(\varphi^{\dag}  i  \overleftrightarrow D_{\mu} \varphi ) (\bar \ell \gamma^{\mu}  \ell)$ and $(\varphi^{\dag}  i  \overleftrightarrow D_{\mu} \varphi ) (\bar e \gamma^{\mu}  e)$ which depend on the kinetic mixing parameter $\kappa$.   This scenario lies outside of the framework assumed in this work, as these operators will contribute to $Z$-decay observables and compete with the loop-induced effects considered here.  

However, if the kinetic gauge mixing parameter vanishes at the matching scale $\Lambda$, we can conclude from our analysis that a viable scenario would be a $Z^{\prime}$ boson with vectorial coupling to muons.   We could then propose a fully-fledged model invoking a $\mathrm{U(1)}^{\prime} = L_{\mu}-L_{\tau}$ gauge symmetry that is spontaneously broken by the vacuum expectation value of a scalar field that is singlet under the SM gauge group.\footnote{The difference of family lepton numbers $L_{\mu}-L_{\tau}$ is anomaly free with the SM fermion content and automatically gives a vectorial $Z^{\prime}$ coupling to muons~\cite{He:1990pn,He:1991qd}.    }     In this case heavy exotic fermions can provide the $Z^{\prime}$ coupling to the top-quark via fermion mixing effects~\cite{Kamenik:2017tnu}.    Having an explicit ultraviolet completion would allow us to calculate one-loop finite corrections to the matching at the high energy scale and test our assumption that the low-energy processes considered are dominated by logarithmic renormalization group evolution induced terms, such task is however beyond the scope of this work.

$\bullet$ It was originally proposed in~\cite{Becirevic:2017jtw} that the scalar LQ $R_2 \sim (3,2,7/6)$ can accommodate $b \to s \ell^+ \ell^-$ anomalies at the one-loop level.      This scenario was also analyzed later in~\cite{Fajfer:2018bfj}.  As we saw,  the LQ $R_2$ only generates the operator $\mathcal{O}_{\ell u}$.    It is worth noting that, as evidenced in Figure~\ref{figplots2}, we can conclude that this scenario has an important tension with LEP-I measurements.    

The model presented in~\cite{Becirevic:2017jtw} reduces this tension slightly by including a coupling of the LQ to the charm quark, besides the coupling to the top-quark.   In this case, there is another relevant contribution to the effective Hamiltonian for $b \to s \ell^+ \ell^-$ transitions from an operator involving both the charm and the top quark $[\C_{\ell u}]_{\mu \mu ct}$.   This contribution can alleviate the tensions between $b \to s \ell^+ \ell^-$ anomalies and LEP-I, however, since this new contribution is suppressed relative to the one from $[\C_{\ell u}]_{\mu \mu tt}$ by a factor $m_c/(m_t V_{tb}  V_{ts}^*)  \sim 1/6$, the required charm coupling of $R_2$ is larger than the top coupling in this model. 

\begin{figure*}[t]
\begin{center}{
\includegraphics[width=7.5cm]{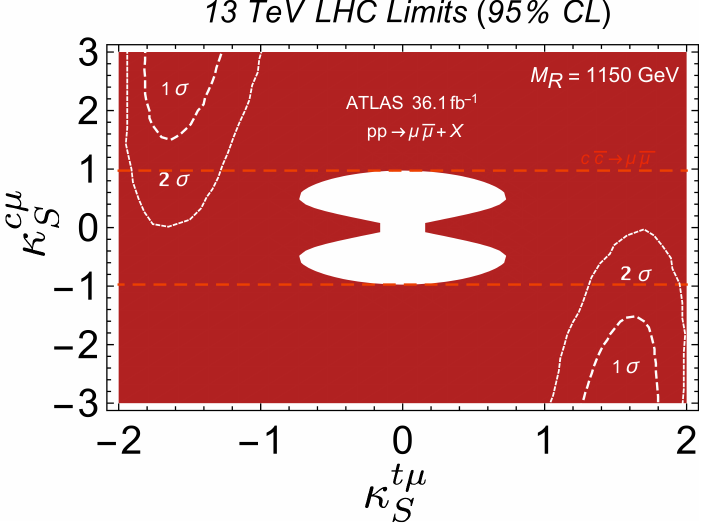}
~
\includegraphics[width=7.5cm]{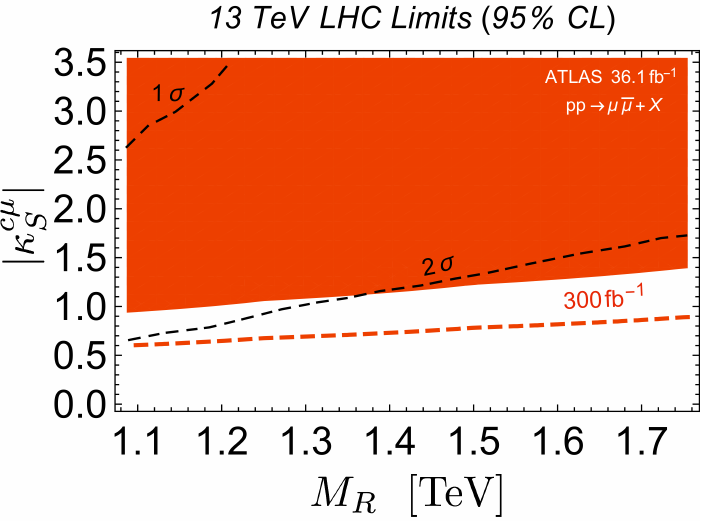} \\
\caption{\textit{Scenario of $R_2$ coupling to charm and top motivated by~\cite{Becirevic:2017jtw}.  Excluded regions by the LHC at $95\%$~CL from a recast of a dimuon search at $36.1$~fb$^{-1}$, including projections to $300$~fb$^{-1}$.   The preferred region by a global fit of $b \to s \ell^+ \ell^-$ and LEP observables at $68\%$~CL and $95\%$~CL is shown by dashed contours.}    \label{figlhcr2}     } }
\end{center}
\end{figure*}

In the following we show that high-$p_T$ searches at the LHC set stringent constraints on this model excluding most of the preferred region by a global fit of $b \to s \ell^+ \ell^-$ and LEP observables. Because of the large $R_2$ coupling to charm, the model predicts a large deviation in the high-$p_T$ di-muon tails at the LHC \cite{Greljo:2017vvb}. For this we recast once again the inclusive $pp \to \mu\bar\mu+X$ search by ATLAS \cite{Aaboud:2017buh} with the NP signal given by the combination of the $t$-channel exchange of $R_2^{5/3}$ in $c\bar c\to \mu\bar\mu$ via the charm-muon Yukawa coupling and pair production of LQs decaying into $R_2^{\dagger} R_2\to t\bar t\mu\bar\mu,\, c\bar t\,(t\bar c)\mu\bar\mu$. We find that the $R_2$ model as an explanation of the $R_{K^{(*)}}$ and  $b \to s \ell^+ \ell^- $ anomalies is excluded for LQ masses below $1.15$~TeV.\footnote{Here we assume all tau-lepton and down-type Yukawa couplings of $R_2$ to be zero while~\cite{Becirevic:2017jtw} does not make this assumption. Additional decay channels of $R_2$ into tau-leptons would reduce the branching fractions for the muonic decay channels making this bound weaker.} In Figure~\ref{figlhcr2} (left) we illustrate this with the dark red exclusion region at 95$\%$ CL for the benchmark $M_R=1.15$~TeV in the Yukawa coupling plane $\{\kappa_S^{t \mu},\kappa_S^{c \mu}\}$, following a notation analogous to \eqref{myEq} for the LQ couplings. The allowed region at $68\%$~CL and $95\%$~CL from a global fit to $b \to s \ell^+ \ell^- $ and LEP-I observables is shown in Figure~\ref{figlhcr2} as dashed contours. The horizontal red dashed contours represent the limit extracted if we had only considered the $t$-channel $c\bar c\to\mu\bar\mu$ in our analysis. Notice that including the final states $t\bar t\mu\bar\mu$ and $c\bar t(t\bar c)\mu\bar\mu$ from pair production in the di-muon recast removes this flat direction in $\kappa_S^{t \mu}$. For a LQ mass above $\sim1.2$~TeV LQ pair production becomes negligible leaving only the $t$-channel mediated process $c\bar c\to\mu\bar\mu$ as the only contribution to the di-muon tails. In Figure~\ref{figlhcr2} (right) we give the 95\% exclusion regions in the $\{M_{R},|\kappa_S^{c \mu}|\}$ plane for this scenario in orange. These bounds only rely on the size of the charm-muon coupling of $R_2$, so they apply to the model in~\cite{Becirevic:2017jtw}.  Between $1.15<M_R< 1.35$~TeV the allowed region at $95\%$~CL from the low energy fit is not completely excluded by this LHC search. Our projections of the di-muon bound to 300\,\invfb~of LHC data, given by the dashed orange contour, cover this last piece of parameter space. 

$\bullet$ As shown in Sec.~\ref{sec:R2U1highpt}, these strong tensions of $R_2$ with current LHC data can be avoided if one trades the dangerous couplings of $R_2$ to charm by a new vector LQ state $\widetilde U_1$ coupling to top. LQ pair production searches, shown in Figure~\ref{fig:massLQs}, put a current lower bound on both masses at about $M_{\widetilde U}\!\sim\!1.9$~TeV and $M_R\!\sim\!1.2$~TeV. For these masses, and for couplings of moderate size, the combination $R_2$ plus $\widetilde U_1$ can successfully explain the $b \to s \ell^+ \ell^-$ anomalies without large tensions with high-$p_T$ and low-energy observables. While the inclusive di-muon searches for LQ pair production is already giving relevant limits on this model, a dedicated search by the LHC for $t\mu$ resonances in $t\bar t\mu\bar\mu$ final states will considerably improve them. In particular, the necessity for large couplings $|\kappa_{S,V}|\gsim2$ to explain the $B$-anomalies singles out the single LQ production mode $pp\to \mathrm{LQ}\, \mu t$ as an additional probe for this model.

It is worth mentioning that one interesting possibility is to consider $\widetilde{U}_1$ as a gauge boson of an $\mathrm{SU(4)}$ gauge extension of the SM. Here, the required non-universal couplings of the vector leptoquark to fermions can be generated with a horizontal gauge group or via mixing with vector-like fermions in an analogous fashion to~\cite{Assad:2017iib,DiLuzio:2017vat,Calibbi:2017qbu}.     

\begin{figure*}[t]
\begin{center}{
\includegraphics[width=7.5cm]{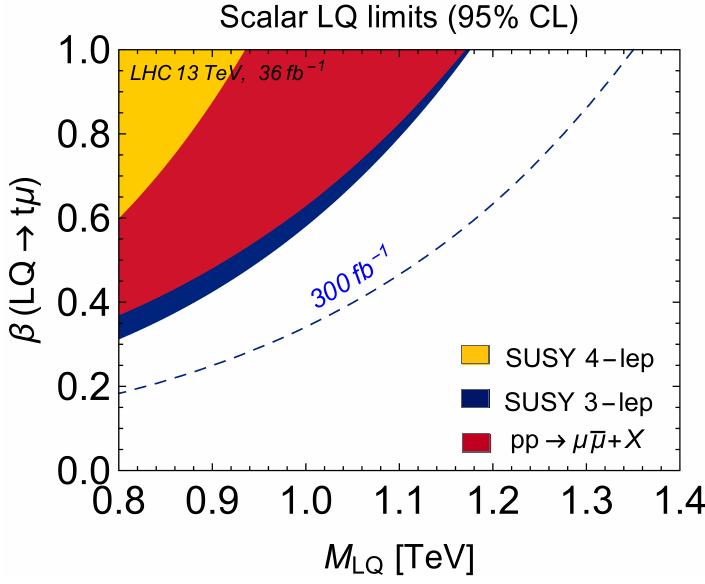}
~
\includegraphics[width=7.5cm]{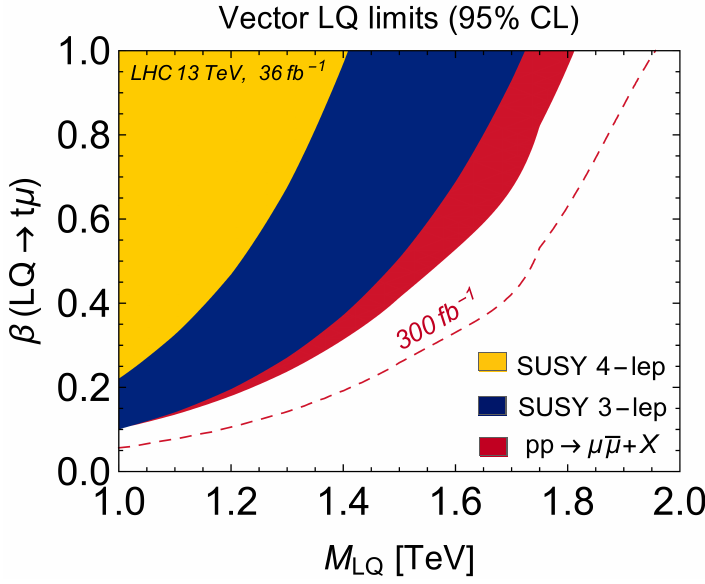}
\caption{\textit{LHC bounds for a pair produced LQ decaying into the $t\bar t\mu\bar\mu$ channel.} \label{fig:LQpairbounds}     } }
\end{center}
\end{figure*}

\section{Conclusions}  \label{seccon}

Current anomalies in $b \to s \ell^+ \ell^-$ transitions could represent the first signature of physics beyond the Standard Model.  Future measurements from the LHCb collaboration should be able to shed light on this possibility in the following years. Furthermore, the Belle-II experiment is also expected to add important information on this subject in the near future.     

In this work we have analyzed a possible explanation of these anomalies with new physics around the TeV scale that couples to right-handed top-quarks and muons.      The required contributions to $b \to s \ell^+ \ell^-$ arise at the one-loop level in this case.   We have explored this scenario both with the Standard Model Effective Field Theory framework as well as with particular models.     We have found a rich complementarity between the flavour observables, high-$p_T$ searches at the LHC, and electroweak precision measurements performed at LEP.  

Considering the two SMEFT operators $\mathcal{O}_{\ell u}= (\bar \ell_{ L \mu} \gamma^{\alpha}  \ell_{ L \mu}) (\bar t_R \gamma_{\alpha}  t_R  )$ and  $\mathcal{O}_{e u}=  (\bar \mu_{R} \gamma^{\alpha}  \mu_{R})  (\bar t_R \gamma_{\alpha}  t_R  )$, we obtain that the preferred Wilson coefficients satisfy $\C_{eu} \sim \C_{\ell u}$, implying that new physics enters in $b \to s \mu^+ \mu^-$ mainly through the Wilson coefficient $\C_9$ of the Weak Effective Theory.  We find that a vector boson in the irreducible representation of the SM gauge group $Z_{\mu}^{\prime} \sim (1,1,0)$ with vectorial coupling to muons and a combination of two leptoquarks, the scalar $R_2 \sim (3,2,7/6)$ and the vector $\widetilde U_{1 \alpha} \sim (3,1,5/3)$ can produce the required new physics pattern. By recasting different new physics searches at the LHC, we showed that high-$p_T$ searches are already probing these mediators in the parameter space region that accommodates the flavour anomalies and exclude an important range of the possible masses. These mediators can therefore be discovered with the increase in luminosity at the LHC.  

Finally, our framework does not explain the anomalies observed in $b \to c \ell \nu$ transitions.    It is interesting to note that, due to the loop suppression of the new physics contribution in $b \to s \ell^+ \ell^-$ within our scenario, we rely on mediators with fermionic couplings of order one and a mass around the TeV scale.    The anomalies in $b \to c \ell \nu$ transitions hint to mediators with these characteristics contributing at tree level, given that it is a tree-level process in the SM.    It can therefore be interesting to extend the framework presented in this work in order to accommodate both $b \to s \ell^+ \ell^-$ and $b \to c \ell \nu$ anomalies, having mediators that enter at the loop and tree level respectively.

\section*{Acknowledgments}

A.C. is supported by the DFG grant  BU 1391/2-1.  D.A.F. is supported by the {\it Young Researchers Programme} of the Slovenian Research Agency under the grant No.~37468.   J. E. C.-M. is supported by the STFC grant ST/P000762/1. D.A.F. would like to thank Jernej F. Kamenik for valuable discussions and for carefully reading the manuscript. The authors would like to dedicate this work to the Venezuelan researchers, professors and students currently struggling to carry out their research and education in Venezuela under inexcusable precarious conditions for which we hold the current government accountable. We would also like to thank the Universidad Sim\'{o}n Bol\'{i}var for the quality education and stimulating atmosphere provided during our undergraduate studies.

\begin{appendix}

\section{Model independent bounds on LQ pair production in the $t\bar t\mu\bar\mu$ channel}  \label{appendixA}

We give results from recasting the $\sim 36$\,\invfb~ SUSY and di-muon tail search \cite{Aaboud:2018zeb,Aaboud:2017dmy,Aaboud:2017buh} for the QCD induced pair production process $pp\to \text{LQ}^\dagger \text{LQ} \to t\bar t\mu\bar\mu$ of a generic LQ state. We present the 95\% CL exclusion limits in Figure~\ref{fig:LQpairbounds} for both scalar and vector LQs with mass $M_{\text{LQ}}$ and branching ratio $\beta(\text{LQ}\to t\mu)$.   The solid lines represent the exclusion bounds from the searches with current luminosity while the dashed lines are for a projected LHC luminosity of $300$\,\invfb. The di-muon tail search produces the most stringent bounds for the vector leptoquark, while for the scalar leptoquark these limits are comparable with those coming from the SUSY tri-lepton search.  
These figures give an indication of how our results get modified  when one allows for additional decay channels for the leptoquarks.

\end{appendix}


\end{document}